\PassOptionsToPackage{table}{xcolor}

\documentclass[sigconf]{acmart}

\usepackage{array, ragged2e}
\usepackage{tabularx}
\usepackage{longtable}
\usepackage{balance}
\usepackage{csquotes}
\usepackage{enumitem}

\hypersetup{
       breaklinks=true,   
       colorlinks=true,   
    }

\newcolumntype{P}[1]{>{\RaggedRight\arraybackslash}p{#1}}

\AtBeginDocument{%
  \providecommand\BibTeX{{%
    \normalfont B\kern-0.5em{\scshape i\kern-0.25em b}\kern-0.8em\TeX}}}
\MakeOuterQuote{"}

\copyrightyear{2023}
\acmYear{2023}
\setcopyright{rightsretained}

\acmConference[SIGCSE 2023] {Proceedings of the 54th ACM Technical Symposium on Computer Science Education V. 1}{March 15--18, 2023}{Toronto, ON, Canada.} 
\acmBooktitle{Proceedings of the 54th ACM Technical Symposium on Computer Science Education V. 1 (SIGCSE 2023), March 15--18, 2023, Toronto, ON, Canada}

\acmDOI{10.1145/3545945.3569846}
\acmISBN{978-1-4503-9431-4/23/03}

\makeatletter
\gdef\@copyrightpermission{
  \begin{minipage}{0.3\columnwidth}
     \href{https://creativecommons.org/licenses/by/4.0/}{\includegraphics[width=0.90\textwidth]{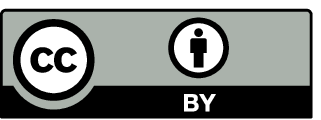}}
  \end{minipage}\hfill
  \begin{minipage}{0.7\columnwidth}
     \href{https://creativecommons.org/licenses/by/4.0/}
     {This work is licensed under a Creative Commons Attribution International 4.0 License.}
  \end{minipage}
  \vspace{5pt}
}
\makeatother

\begin{document}

 \fancyhead{}
 
\title{"This Applies to the Real World": Student Perspectives on Integrating Ethics into a Computer Science Assignment}

\author{Julie Jarzemsky}
\email{julie.jarzemsky@colorado.edu}
\affiliation{%
  \institution{University of Colorado Boulder}
}

\author{Joshua Paup}
\email{joshua.paup@colorado.edu}
\affiliation{%
  \institution{University of Colorado Boulder}
}

\author{Casey Fiesler}
\email{casey.fiesler@colorado.edu}
\affiliation{%
  \institution{University of Colorado Boulder}
}

\renewcommand{\shortauthors}{Julie Jarzemsky, Joshua Paup, \& Casey Fiesler}

\begin{abstract}
  There is a growing movement in undergraduate computer science (CS) programs to embed ethics across CS classes rather than relying solely on standalone ethics courses. One strategy is creating assignments that encourage students to reflect on ethical issues inherent to the code they write. Building off prior work  that has surveyed students after doing such assignments in class, we conducted focus groups with students who reviewed a new introductory ethics-based CS assignment. In this experience report, we present a case study describing our process of designing an ethics-based assignment and proposing the assignment to students for feedback. Participants in our focus groups not only shared feedback on the assignment, but also on the integration of ethics into coding assignments in general, revealing the benefits and challenges of this work from a student perspective. We also generated novel ethics-oriented assignment concepts alongside students. Deriving from tech controversies that participants felt most affected by, we created a bank of ideas as a starting point for further curriculum development.
\end{abstract}


\keywords{ethics, introductory programming, CS1, social impact, assignments, university, undergraduate, content, focus groups, content moderation}

\maketitle

\section{Introduction}
Rising awareness of the harm and injustice that technology can inflict upon people and society has led to calls for increased critique and interrogation of current practices within the tech industry \cite{Benjamin2019race,Costanza2020design,Oneil2016weapons}, as well as proposals to expand the inclusion of concepts such as ethics, social impact, and justice within computer science and other technology-related education programs \cite{Lin2022cs, Vakil2018ethics,Fiesler2020we,Ko2020}.  Though in the U.S., the Accreditation Board for Engineering and Technology requires computer science curricula to educate about the impacts of computing \cite{Abet2021}. Solutions for accomplishing this vary. For example, many undergraduate CS programs satisfy this requirement with standalone ethics courses \cite{Homkes2009meeting}, which though an important component of ethics education \cite{Fiesler2020we}, have the drawback of being potentially disconnected from technical content. 

In addition to these standalone ethics classes, there is also now a growing movement within universities to embed ethics across the entire CS curriculum \cite{CaliffGoodwin2005, grosz2019embedded,Mozilla2021}, with one common strategy being incorporating ethical and social context into technical assignments. Recent examples from prior work have situated assignments in the context of ethical dilemmas such as ad personalization, college admissions algorithms, bias in criminal justice systems, and privacy \cite{CaliffGoodwin2005, Fiesler2021integrating, Davis2011incorporating, Ryoo2013democratizing, Kesar2016including}. These efforts also go beyond learning about ethics in general to instilling a responsibility within CS students to consider the impacts of the code they write. This is a step towards what Amy Ko calls a “critical literacy of computing,” beyond “just an ethics requirement for CS majors…recasting computing itself in moral, ethical, and social terms” \cite{Ko2020}. 

To contribute to these efforts, we created a programming assignment for an introductory CS class centered on the topic of automated content moderation. After creating the initial version of this assignment, we conducted focus groups with students at our university to gather detailed feedback on the assignment and the ethical discussions it prompted. The focus groups enabled us to expand on previous work that surveyed introductory computing students on their attitudes towards ethics based assignments \cite{Fiesler2021integrating}, in order to gather more detailed feedback that both provided guidance for iterating on the assignment, and ideas that will help guide the integration of ethics into CS curriculum in a way that centers student engagement.

\section{Creating the Assignment}
We began our process by reviewing literature and existing ethics-based assignments, discussing assignment concepts with colleagues, and creating a list of social issues that the assignment could focus on. We ultimately decided to focus on automated content moderation, given its abundance of both related technical skills and ethical considerations, as well as the authors’ own observations of discussions in a standalone ethics class that were highly engaging to students. We noted that much of the literature about the technical challenges of automated content moderation involve issues such as hate speech detection \cite{Sap2019social, Vincent2016twitter}, but due to the obvious potential for harm we did not want to ask students to write code that directly involved hate speech. We instead considered the shape of the problem in terms of automated language detection and bad actors, and created a hypothetical problem: writing and reasoning about a content moderation system for a cat-run social media platform. The following context was provided for the assignment:

\blockquote{Catter is a social media platform built by and for cats.  The cats’ platform has recently been getting spammed by dogs, so they have decided to remove all mentions of dogs from their platform entirely. However, cats are not great at programming.  They need your help in removing all of the dog content from their platform.}

We used Python for our assignment since this is a common language for intro-level courses. The assignment covers file input and output, parsing text using regular expressions, functions, loops, conditionals, and using a natural language processing library (Python’s NLTK library).
 
The assignment asks students to write several iterations of a content moderator for Catter. First, students must simply detect whether a string has the word “dog” in it, scan a file to remove all posts with occurrences of “dog,” and write the filtered content to a new file. In the next section, the dogs in our hypothetical scenario learn to bypass this filter, and the students are asked to update their system to detect a list of words, including slang such as “doggos” and “dawg.” They are asked to read in a file that contains this list, then use their previous code to remove all posts with those words. Finally, the dogs start posting negative comments about cats, bringing in the need for the NLTK sentiment analyzer. The last part requires the students to remove any posts containing the word “cat” that the analyzer marks as having a negative sentiment. These different steps are designed to illustrate challenges a real developer might face when moderating content with code: wrestling with users who “outsmart” the algorithm so their content gets posted--regardless of whether they are bad actors or are themselves wrestling with systematic errors in the system.

In order to encourage students to consider the nuances within content moderation beyond this hypothetical scenario, and to examine the decisions they made in their own code, we followed the coding exercise with a reflection section. In this part of the assignment, students watch Maarten Sap’s talk on Social Bias Frames, which shows how binary keyword matching, like the students were required to code, can miss sexist, racist or otherwise hateful speech \cite{Sap2020talk}. The assignment asks for written responses to questions that prompt students to critique the way that the code they wrote moderates content. We chose to include written reflections because previous work suggests that integrating reflections is critical in encouraging students to consider their ethical own responsibility in creating technology \cite{Brown2022shortest}. The reflection questions also broaden the discussion, asking students to provide a news article related to the topic of content moderation. Our aim here was to expand on real world connections, which has been shown to have a positive effect on student engagement \cite{Guzdial2009education, Nickerson2015grounding}, including increasing participation amongst underrepresented groups in computing \cite{Khan2016computing}.

\section{Evaluation Methods}
To evaluate the design of the assignment, we created a plan for 90-minute focus groups with 3-5 participants each, all of whom had already taken an intro programming course. These groups participated in the following tasks:
\begin{enumerate}
    \item Pre-questionnaire about content moderation
    \item Review of coding portion of assignment
    \item Group discussion and individual surveys on coding portion:
    \begin{enumerate}
        \item Reactions to the assignment
        \item How they would complete it
        \item Any changes they would suggest
    \end{enumerate}
    \item Review of reflection portion
    \item Group discussion and individual surveys on reflection portion
    \item Brainstorm new assignment ideas centered around ethical dilemmas
    \begin{enumerate}
        \item Generated list of tech controversies
        \item Discussed how these controversies tie in with CS topics
    \end{enumerate}
\end{enumerate}

One key goal for the group discussion and questionnaires was to gauge whether and how this assignment made participants consider ethical dilemmas within content moderation, through the coding alone or with the supplementary reflection. For the final brainstorming session where the students thought about how we might integrate ethics into programming assignments more broadly, we drew from our own assignment creation process, starting with listing social issues the assignment could be based on. We created the list by asking participants “What controversies or impacts of technology have you either heard about or been personally affected by?” which they responded to by posting on a shared Google Jamboard. Afterwards, we asked participants to choose one controversy that impacts them or other students the most. Finally, we presented the students with a list of CS topics and asked how they might tie in with the list of tech controversies they created.

We ran one pilot focus group with members of our lab, and after the study was approved by the university’s institutional review board, began recruiting participants, looking for anyone who had taken an introductory programming course at our large, predominantly white, public university in the United States. The recruitment materials, which were shared with computer science professors and through flyers on campus, mentioned only that participants would be evaluating CS curriculum. We did not mention that the assignment involved ethics to reduce selection bias of students who had a specific interest in ethics.

We ultimately had a total of 16 individual participants, with 2-5 participants per group. Our participants were majority white, with a roughly even split between men and women. The focus groups included the following participants:
\begin{itemize}
    \item Group 1 (P1-3): a freshman, sophomore, and junior, all CS majors
    \item Group 2 (P4-7): a freshman CS major, two juniors with dual majors in CS and applied math and economics, and a graduate student CS major
    \item Group 3 (P8-10): a junior and sophomore information science majors, and a senior with dual majors in CS and mechanical engineering
    \item Group 4 (P11-12): two graduate student CS majors
    \item Group 5 (P13-16): a freshman aerospace engineering major, a senior CS major, a graduate student information science major,and  a graduate student CS major
\end{itemize}

We adapted the plan as we ran the groups, shortening the section about how the assignment would be solved in exchange for more time for reflecting on how the assignment prompted ethical reasoning and on integrating ethics in general.

Following the focus groups, we conducted a thematic analysis of the transcripts \cite{Clarke2015thematic}. Two of the authors highlighted anything of note in the transcripts and created a list of themes from these highlights. We then performed a second round of coding for those specific themes and finally we wrote memos about each theme, which we synthesized into our findings and discussion. During this process, all authors met to discuss and iterate on themes. Our findings below represent the synthesis of these themes, including representative quotes.

\section{Findings}
From the focus groups, we gained perspective on the strengths and weaknesses of the assignment and for student opinions integrating ethics in general. We also created a bank of new CS assignment ideas integrated with ethical topics that the students in our groups cared about.

\subsection{Strengths of the assignment}
Using a hypothetical scenario within the assignment was successful from our observations and from our participants' perspectives. Starting with the scenario of a social media site for cats neutralized the subject of content moderation, which can include politically or emotionally charged subjects such as misinformation and hate speech. However, the hypothetical still promoted discussion on a topic that can be uncomfortable for students, as well as educators, to talk about, by allowing them to bring up these issues themselves. Participants responded in the questionnaires that the pretend scenario was helpful for easing into reflection on content moderation before jumping into controversial real world issues. The groups readily connected the fake scenario to real issues, with the benefit that students conjured up these issues organically, instead of the assignment feeding them specific situations.

The assignment brought to mind real-world issues in content moderation, even prior to reviewing Sap’s video and the reflection questions. When asked if the coding piece made participants think differently about content moderation, some responded that it did, bringing up issues like censorship, Facebook filtering out content, bias in automated systems, and misinformation. In group discussion, when asked what the implications of the cats’ moderator are, participants immediately tied the hypothetical situation to reality. Many thought the content moderator resembled “echo chambers” within real social media platforms, which limit content to only include opinions the user agrees with. Another theme discussed across groups was over-moderation of content.  One participant gave the example of YouTube removing benign comments, which negatively affects both the content creators and the commenters. They also reasoned about the fairness of moderation systems defined by companies, with P16 reflecting that removal of content is “\textit{not within the users’ power, the power is in the platform,}” and another participant responding that “\textit{the gatekeepers are often a small group of executives. There aren't many regulations that companies need to follow.}”

The assignment caused P14 to reconsider third-party tools she used in other courses, saying that the assignment “\textit{shows you that you might be given a function, or a tool, but how much do you just put your blind trust in it? In my class, we were given a lot of like functions that we would have to put together into our program but that we didn't write ourselves. But [it] makes you wonder ‘how was this written? How does it work?’ even if it's supposed to be a black box type thing…you can't just go and completely put your trust in like the sentiment analyzer or the AI program that's supposed to be moderating it.}” For them, the assignment spurred critical ethical thinking about their coding.

All groups were able to create a feasible plan for completing the code within a short time-frame. Most participants agreed that the assignment would be appropriate for an intro-level course, but some gauged its difficulty as mid or even high-level. The next section outlines modifications we could make to the assignment to address the difficulty level, as well as other improvements participants suggested.

\subsection{Suggestions for improving the assignment}
To reduce the difficulty of the assignment to an introductory level, participants suggested alternatives to regular expressions for the keyword matching or ways to provide more support, such as code examples using regular expressions for students to work off of. Participants also thought installing the NLTK library and understanding sentiment analysis would be difficult for intro-level students. The sentiment analysis requirement could be removed from the coding section but be kept as a discussion topic within the reflection. Depending on the difficulty level and other aspects of the course, making these kinds of adjustments could be critical; prior work has noted that for programming assignments that integrate ethics, when the students are struggling too much with the technical content, they focus less on the contextual aspects \cite{Fiesler2021integrating}.

Participants also offered ideas for additions to the coding requirements. P10 proposed requiring students to give each other example phrases that could break their partner’s moderator, and then build something to handle those phrases. This method, which is known as “Build it, Break it, Fix it” within cybersecurity competitions, has been used in other ethics-oriented technical assignments \cite{Hod2022data}. P16 recommended creating a display of phrases that were moderated out, and creating a tagging system to categorize moderated phrases.  Research on empowering real creators over moderation of their content has shown demand for features like this \cite{Jhaver2022designing}. These ideas could be added to a follow-up assignment or to modify the existing assignment for a higher-level course.

Participants also provided constructive feedback about how students given this assignment would reflect on the ethics of content moderation. We observed from our analysis that discussion was heavily biased towards the negative effects of content moderation with little discussion of cases where it could be beneficial. One topic rarely mentioned within the focus groups was the impact of false negatives, when harmful phrases pass through a moderator undetected. This may be partially due to participants’ pre-existing beliefs about content moderation, but the scenario in the assignment could be adjusted to show how a lack of moderation can have negative impacts just as over-moderation can. For example, the assignment could include examples of posts that are reasonable to moderate out, rather than the cats’ current strategy leaning towards censorship. P15 recommended incorporating fake facts or news that dogs have spread, like providing false information about a vet clinic to steer cats away from getting health care. Based on this feedback we were able to make a number of improvements on the assignment which we will cover in more detail after the findings.

\subsection{Opinions on integrating ethics}
Aside from gathering feedback on this specific assignment through the focus groups, we also aimed to gauge students’ opinions of integrating ethical concepts in general. Their reactions to including an ethical reflection within this assignment were mostly positive (though we acknowledge the possibility of response bias within the study, given the topic). All but two participants responded in the questionnaire that they think Sap’s video and the reflection belong within a CS class. They thought it would prompt students to think about the direct impacts of technology they are contributing to, something “\textit{many students up until this assignment may never have been challenged about,}” according to one participant.  They stressed the importance of reflecting on impacts, one writing: “Implications of algorithms and ethics should be part of CS curriculum. People creating the problems should also be fixing them or at least know about them.”

Participants also mentioned the potential of the real world context to spark interest in more advanced topics. It “\textit{makes students think more about applications out in the real world…maybe if people are interested, you could point them to the class for natural language processing, [for] down the road,}” one remarked.  Another responded that “contextualizing” the assignment “provides extra motivation/understanding and maybe career insight,” suggesting that including a context could motivate students to continue pursuing careers within technology.

However, participants did raise concerns about the integration of these topics into technical assignments. Some expressed that they would rather study ethics outside of CS curriculum, and prefer it be in a standalone course. “\textit{If I am expecting and wanting a class to be technical in nature, I would likely be unhappy writing a paper since it is not what I signed up for,}” P6 explained. Others mentioned they would see the reflection portion of the assignment as a “\textit{grade bump.}” P8, who had previously experienced ethics content in an economics course, said they did not have to reason deeply about conflicts to complete the assignment. It is a challenge to determine how to grade whether a student has really thought through the impacts of the technology in question. Provoking meaningful thinking while not adding too much additional work is a fine line to tread.

Students are already working hard to learn the technical skills of programming, so adding extra work could become a burden rather than an opportunity to reflect and learn. For our assignment specifically, building a fair moderation system seemed too difficult to accomplish for some participants. Many expressed that real-world content moderation systems are so flawed that we are better off with no moderation at all. This presents a challenge for integrating ethics: how can we incorporate ethics without only focusing on discussions of harmful technologies, but also create solutions to these difficult problems?

Another concern a small number of participants raised was that ethics-oriented assignments carry the risk of inappropriately imposing a particular set of values on students. “\textit{Make the students think about this on their own, rather than trying to influence their opinions,}” suggested one student. Another raised concerns about conformity; that students would side with the opinion that the majority of the class was expressing rather than speak their own mind. Ethics can be a thorny subject, and uncharted territory for some CS instructors, but the effort to incorporate them is worth it. We explain suggestions for facing these barriers in the discussion.

\subsection{New assignment ideas}
One other product of the groups was ideas for new ethics-based assignments. Collaborating with students proved to be a fruitful method for creating assignment ideas. Participants generated a list of tech controversies they felt most affected by, then connected these controversies with assignment ideas and corresponding concepts covered in CS curriculum. Several of these assignment ideas from our participants can be found in \autoref{tab:Table 1}. 
\begin{table}
    \centering
    \caption{The first column represents a controversy in the tech industry that our participants identified, followed by an assignment idea that could be developed around that controversy in the second column, and the assignment's corresponding CS topic in the third column.}
    \resizebox{\linewidth}{!}{%
    \begin{tabular}{|P{0.238\linewidth}|P{0.435\linewidth}|P{0.265\linewidth}|}
        \hline
        \textbf{Context} & \textbf{Assignment Idea} & \textbf{CS Topic} \\ 
        \hline
        Algorithms for organ donor matching & Sort recipients using different algorithms and discuss trade-offs between them & Sorting algorithms \\ 
        \hline
        Addiction to social media & Create a program that decides how social media posts are ordered & Conditionals, trees, graphs \\ 
        \hline
        Addiction to social media & Design UI/UX that increases or decreases screen time & UI/UX design \\ 
        \hline
        Dark patterns on the web & Design a system to make it difficult or easy to unsubscribe from an email list or opt out of cookies & UI/UX design \\ 
        \hline
        Influence of search engines on what news stories users see & Use various search or sort algorithms to create a list of ordered results of news stories & Sorting algorithms, search algorithms, object-oriented programming \\ 
        \hline
        Misuse of user data and right to be forgotten & Compare data storage methods and the complexities of keeping data private and ensuring complete deletion & Memory allocation and management, object-oriented programming \\ 
        \hline
        User data and privacy & Write a program that infers information about a user based on existing user data                                     & Conditionals, variable \\ 
        \hline
        Usability and inclusivity in web forms & Compare from different users’ perspectives how inclusive different web forms are in terms of race, gender, sexuality & UI/UX design, personas \\ 
        \hline
        Mental health impacts of social media apps & Create an image filter that changes the sentiment of the image & Image processing, could use machine learning \\
        \hline
    \end{tabular}
    }
    \label{tab:Table 1}
\end{table}

One group agreed that the worst of the tech controversies was addiction to social media and “doom-scrolling,” when a user gets stuck endlessly scrolling through content. They proposed a sorting algorithm assignment, P15 saying “\textit{you are recommended content on TikTok based off of how relevant it is to you, but also how popular it is, right? So…how do you decide what is sorted to the top?}” Another group suggested an assignment to create a user interface or experience to either increase or decrease screen time.

Others situated sorting algorithms in the context of search results for news, and how this could influence a users’ opinions over time. P8 suggested graph algorithms for determining relevance, drawing from an example she had seen in another course, where she determined sort order based on how many connections a node had to other nodes in a graph.

Another idea came from a blood type matching algorithm P7 was required to write for an algorithms class she had taken.  She thought it could be expanded to have ethical reasoning about organ donor matching. The algorithm could “\textit{take into account how long someone's been on the organ waiting list, how fresh the organ is, and then assign it based on age},” according to this participant. We could see a scenario where students create different versions of the system and run tests for various groups, discussing the trade-offs in fairness for each system.

Another common controversy was data collection and privacy. Participants and the first author discussed one assignment idea to develop an object-oriented design or a database representing user data that highlights how much information platforms gather on their users. The assignment could incorporate discussions of how engineers manage data, especially in cases where it needs to be erased for privacy reasons.  Participants also mentioned the process of data extraction. P2 remarked that a system “\textit{could have your favorite TV show be known just by what you've been clicking…your race, gender, and name}.” Perhaps students could develop and reason about systems that infer information about a user based on pieces of data they have collected. Targeted ads also came up, a subject which has already been incorporated into CS assignments by other researchers \cite{Fiesler2021integrating}.

\section{Improvements to the assignment}
In considering both the specific feedback from participants and their reflections about this type of assignment in general, we iterated on our original assignment. After observing the participants think through how they would complete the coding portion of the assignment, we modified the wording of the instructions to make them easier to read. Instead of using a paragraph of text to lay out instructions, we detailed out steps in a list and made it more clear what each function in the skeleton code should do. Because many participants noted the NLTK toolkit as the most challenging part of the assignment, and that it would be useful to have more background information, we added a more detailed explanation of the output and a link to a guide on using this library. We also removed a question from the reflection section about comparing binary to structured toxic speech detection that some participants found difficult to answer given the 3-minute video.

We also received suggestions for extending the assignment, and included these in the resources we are providing to instructors who want to use the assignment. These extensions included adding a tagging system to moderate posts, allowing users to input certain words they would like to moderate, creating a display of posts that were flagged for removal, and including a pair programming activity where students would try and break each others’ moderation systems, then modify code to solve for those cases.

In addition to the changes we implemented, we have ideas for other adaptations to the assignment. The feedback indicated the need to balance out the scenario so that students understand the potential benefits of content moderation. The participants gave us specific suggestions on how we might do this: by including examples where the dogs were posting misinformation that could be detrimental to the cats’ lives, for example. For the reflection, the example detailed in the video is offensive to women. Students could be asked to provide examples of how other groups are affected by false negatives or positives in content moderation. Scholarly works about bias against marginalized groups in speech toxicity detection systems could also be discussed \cite{Sap2019risk, Thiago2021fighting,Haimson2021disproportionate}. Instructors using this assignment could consider using in-class discussion rather than written responses for the reflection piece of the assignment. This would address students’ aversion to being required to do written work within a technical assignment. We also observed benefits in the focus group of having a group discussion. The group discussion allowed participants to hear each others’ viewpoints instead of reflecting individually. Opting for a discussion would also take out the need for class staff to grade written responses.

\section{Discussion}
We believe CS educators can empower students to become more proactive in considering social impacts of their work. Embedding ethics topics and discussions throughout computer science coursework can help accomplish this, and has other positive side effects. Including ethics-based assignments can bolster cultural competency when it brings up issues such as racial or gender bias in AI, preparing students to be more inclusive and equitable when they enter an industry known for the opposite \cite{washington2020twice}. Previous work also shows that including a real-world context can increase retention of underrepresented groups in computing \cite{Khan2016computing}. Our participants reported our assignment to be more interesting and memorable than a standard programming assignment, reinforcing findings from previous work that integrating ethics improves engagement \cite{Fiesler2021integrating}.

Despite these benefits, we also recognize that embedding ethics can be daunting. In this section, we give recommendations in creating ethics-based assignments. We explain why gathering feedback was useful for us, give strategies for tackling controversial subject matter, and measure the trade-offs of hypothetical vs. real scenarios for sensitive subjects.

Through our focus groups, we tested the effectiveness of embedding ethics within our new CS assignment. The participants’ feedback positively reinforced our methods of embedding ethics and gave us ideas for improving the assignment. The feedback helped us gauge whether including a reflection in addition to embedding the coding situation in an ethical scenario was necessary. We found that the reflection was worth including. 

Prior to the reflection part of our assignment, few participants acknowledged the potential benefits of content moderation or considered the impacts of false negatives, when content that should be removed passes through a moderation system undetected. The discussions in the focus groups revealed that the scenario in our coding portion highlighted the impacts of false positives, when benign content is flagged for removal, and failed to draw attention to false negatives. This may also be due to participants’ experiences with content moderation outside of the focus groups, since both false positives and false negatives in content moderation often disproportionately impact people from marginalized groups, such as people of color \cite{Haimson2021disproportionate, Ohlheiser2021welcome}. Participants in our groups, which reflected the population of our predominantly white university, may have never personally experienced the negative effects of false negatives. In a study of an ethics-related assignment, Klassen and Fiesler found that when speculating about ethics in the classroom, students and instructors may fail to consider perspectives outside of their own; as an instructor in their study said, “People tend to lean on their own experiences pretty heavily in speculation, and don’t, unless they’re very carefully prompted, consider broader context” \cite{Klassen2022run}. Including a reflection is an opportunity to “carefully prompt” students to consider other perspectives.

In the focus groups we were also able to test out using a hypothetical situation, which we found to be effective. However, instructors do need to be careful about their transitions to real issues from the hypothetical. In our case, at least one participant found the transition from the cats’ platform into more charged topics like racist hate speech “too sudden.” There is also the risk that students will fail to bridge the gap between the hypothetical situation and the severity of the ethical impact in real life. While we did not observe this to be the case for our assignment, we believe it’s important to ensure that students are not getting a watered-down representation of the ethical issue being addressed by an assignment. We found the benefits of using a hypothetical scenario worth these extra considerations, because this approach removed barriers in considering ethics for those averse to controversial discussions.

Running the focus groups provided us with valuable feedback and we recommend gathering feedback to anyone creating ethics-based CS assignments. Without hearing from our participants, we would have remained unaware of the blind spots we had to issues with our assignment. Some of the improvements listed above came directly from students in our groups. We got suggestions for confronting barriers they perceived students would have with ethics-based content. We also generated a wealth of ideas for more ethics-oriented assignments, spending only 30 minutes or less in each of the five focus groups. The tactic we used to facilitate brainstorming, starting from tech controversies and then drawing connections to CS topics, flowed more easily than simply asking students if they had ideas for ethics-based CS assignments. The engagement in the brainstorming activity was also encouraging; participants were active in generating ideas and applying them to CS topics.

Once we had student feedback showing their concerns with ethics-based content, we could create some mitigation strategies. Some participants shared concerns about ethics-oriented assignments imposing a particular belief onto them about a particular technology. One article on integrating ethics states that “a good technology ethics course teaches students how to think, not what to think, about their role in the development and deployment of technology.” \cite{Burton2018teach}. Consider the whole picture of any ethical context you integrate and keep questions posed to students open-ended. We made edits along these lines to our focus group script after piloting it with a group from our lab. In place of “Do you think that the moderation system you implemented is harmful?” we asked “What are the implications of the moderation system?”.

Finally, we strongly encourage CS educators to utilize the resources around them when designing similar assignments. Our content moderator assignment materials are publicly available.\footnote{\href {https://www.internetruleslab.com/ethicsbased-computer-science-assignments\#content-mod}{https://www.internetruleslab.com/ethicsbased-computer-science-assignments\#content-mod}} There are also many open source ethics-based CS assignments created by other educators, such as those developed as part of Mozilla’s Responsible Computing Challenge \cite{Mozilla2021}. We also invite educators to use the ideas generated by our participants to create new assignments that fit into their courses. Educators looking to test their assignments could use teaching assistants to pilot assignments, therefore making them collaborators in the process as well, in place of running more time-consuming focus groups. Providing in-depth feedback could also be presented as an extra-credit opportunity for students. Incorporating ethics may seem daunting, but can be accomplished, as others have suggested, by starting small and collaborating between disciplines, adapting existing assignments to include an ethics-oriented context \cite{Fiesler2021integrating}.

\section{Conclusion}
Rather than relying on a standalone course for teaching the ethical implications of the technology students will produce, there is a growing movement to embed ethics throughout the entire computer science curriculum. In this experience report, we demoed an assignment to five small groups of participants that reflected on the design and implications of a content moderation system for a cat-driven social media platform. Discussions with these participants revealed students’ opinions about integrating ethics into coursework, feedback on the assignment itself, and ideas for future assignments. With these findings in mind, we recommend that CS educators creating ethics-based assignments make use of student or TA feedback to improve assignments, use strategies to limit imposition of biases on the ethical dilemma, consider using hypothetical scenarios in their assignments, and embrace resources around them. The benefits of integrating ethics into CS curriculum, including better student engagement as well as preparing them to create technology with societal impacts in mind, are worth the effort of creating ethics-based assignments.

\section{Acknowledgments}
This research was supported by Omidyar Network as well as the Responsible Computer Science Challenge (with funding from Mozilla, Omidyar Network, Schmidt Futures, and Craig Newmark Philanthropies). We would like to thank our participants and the Internet Rules Lab for their feedback on this project, particularly Ella Sarder, Johnny Sreenan, and Camryn Kelley.

\bibliographystyle{ACM-Reference-Format}
\balance
\bibliography{sample-sigconf}

\end{document}